\documentclass{emulateapj}

\usepackage{amsmath,amssymb}

\newcommand{\mb}{\mathbf}
\newcommand{\gb}{\boldsymbol}
\newcommand{\icarus}{Icarus}
\newcommand{\jchemphys}{J.~Chem.~Phys.}

\newcommand{\annstat}{Ann.~Stat.}

\newcommand{\shortlowWav}{290}
\newcommand{\shortuppWav}{450}
\newcommand{\shortcenWav}{413}
\newcommand{\longlowWav}{450}
\newcommand{\longuppWav}{570}
\newcommand{\longcenWav}{510}
\newcommand{\binAlowWav}{290}
\newcommand{\binAuppWav}{340}
\newcommand{\binAcenWav}{325}
\newcommand{\binBlowWav}{340}
\newcommand{\binBuppWav}{390}
\newcommand{\binBcenWav}{368}
\newcommand{\binClowWav}{390}
\newcommand{\binCuppWav}{435}
\newcommand{\binCcenWav}{416}
\newcommand{\binDlowWav}{435}
\newcommand{\binDuppWav}{480}
\newcommand{\binDcenWav}{459}
\newcommand{\binElowWav}{480}
\newcommand{\binEuppWav}{525}
\newcommand{\binEcenWav}{502}
\newcommand{\binFlowWav}{525}
\newcommand{\binFuppWav}{570}
\newcommand{\binFcenWav}{547}

\newcommand{\shortmedDelta}{126}
\newcommand{\shortlowDelta}{36}
\newcommand{\shortuppDelta}{37}

\newcommand{\longmedDelta}{1}
\newcommand{\longlowDelta}{30}
\newcommand{\longuppDelta}{37}

\newcommand{\binAmedDelta}{142}
\newcommand{\binAlowDelta}{175}
\newcommand{\binAuppDelta}{176}

\newcommand{\binBmedDelta}{123}
\newcommand{\binBlowDelta}{87}
\newcommand{\binBuppDelta}{86}

\newcommand{\binCmedDelta}{102}
\newcommand{\binClowDelta}{48}
\newcommand{\binCuppDelta}{48}

\newcommand{\binDmedDelta}{53}
\newcommand{\binDlowDelta}{36}
\newcommand{\binDuppDelta}{37}

\newcommand{\binEmedDelta}{-35}
\newcommand{\binElowDelta}{36}
\newcommand{\binEuppDelta}{45}

\newcommand{\binFmedDelta}{7}
\newcommand{\binFlowDelta}{36}
\newcommand{\binFuppDelta}{43}

\newcommand{\shortmedAg}{0.40}
\newcommand{\shortlowAg}{0.11}
\newcommand{\shortuppAg}{0.12}
\newcommand{\longmedAg}{0.00}
\newcommand{\longlowAg}{0.10}
\newcommand{\longuppAg}{0.12}
\newcommand{\binAmedAg}{0.45}
\newcommand{\binAlowAg}{0.55}
\newcommand{\binAuppAg}{0.55}
\newcommand{\binBmedAg}{0.39}
\newcommand{\binBlowAg}{0.27}
\newcommand{\binBuppAg}{0.27}
\newcommand{\binCmedAg}{0.32}
\newcommand{\binClowAg}{0.15}
\newcommand{\binCuppAg}{0.15}
\newcommand{\binDmedAg}{0.17}
\newcommand{\binDlowAg}{0.11}
\newcommand{\binDuppAg}{0.12}
\newcommand{\binEmedAg}{-0.11}
\newcommand{\binElowAg}{0.11}
\newcommand{\binEuppAg}{0.14}
\newcommand{\binFmedAg}{0.02}
\newcommand{\binFlowAg}{0.12}
\newcommand{\binFuppAg}{0.14}

\shorttitle{The deep blue color of HD\,189733\lowercase{b}: albedo measurements with \emph{HST}/STIS at visible wavelengths}
\shortauthors{Evans et al.}

\begin{document}

\title{The deep blue color of HD\,189733\lowercase{b}: albedo measurements with \emph{HST}/STIS at visible wavelengths}

\author{Thomas~M.~Evans\altaffilmark{1}, Fr\'{e}d\'{e}ric.~Pont\altaffilmark{2}, David~K.~Sing\altaffilmark{2}, Suzanne~Aigrain\altaffilmark{1}, Joanna~K.~Barstow\altaffilmark{1}, Jean-Michel~D\'{e}sert\altaffilmark{3,7}, Neale~Gibson\altaffilmark{4}, Kevin~Heng\altaffilmark{5}, Heather~A.~Knutson\altaffilmark{3}, Alain~Lecavelier des Etangs\altaffilmark{6}}

\altaffiltext{1}{ Department of Physics, University of Oxford, Denys Wilkinson Building, Keble Road, Oxford OX1 3RH, UK }
\altaffiltext{2}{ School of Physics, University of Exeter, EX4 4QL Exeter, UK }
\altaffiltext{3}{ Division of Geological and Planetary Sciences, California Institute of Technology, Pasadena, CA 91125, USA }
\altaffiltext{4}{ European Southern Observatory, Karl-Schwarzschild-Strasse 2, D-85748 Garching, Germany }
\altaffiltext{5}{ University of Bern, Center for Space and Habitability, Sidlerstrasse 5, CH-3012 Bern, Switzerland }
\altaffiltext{6}{ Institut d’Astrophysique de Paris, UMR7095 CNRS, Universite Pierre et Marie Curie, 98 bis Boulevard Arago, F-75014 Paris, France }
\altaffiltext{7}{ Sagan Postdoctoral Fellow }

\slugcomment{To appear in The Astrophysical Journal Letters; received 2013 May 31; accepted 2013 June 15}

\begin{abstract}
We present a secondary eclipse observation for the hot Jupiter HD\,189733b across the wavelength range \shortlowWav--\longuppWav\,nm made using the Space Telescope Imaging Spectrograph on the \emph{Hubble Space Telescope}. We measure geometric albedos of $A_g = \shortmedAg \pm \shortuppAg$ across \shortlowWav--\shortuppWav\,nm and $A_g < \longuppAg$ across \longlowWav--\longuppWav\,nm at $1\sigma$ confidence. The albedo decrease toward longer wavelengths is also apparent when using six wavelength bins over the same wavelength range. This can be interpreted as evidence for optically thick reflective clouds on the dayside hemisphere with sodium absorption suppressing the scattered light signal beyond $\sim$\longlowWav\,nm. Our best-fit albedo values imply that HD\,189733b would appear a deep blue color at visible wavelengths.
\end{abstract}


\section{ Introduction } \label{sec:intro}

The wavelength-dependent manner in which a planetary atmosphere reflects incident starlight reveals valuable details about its structure and composition. In this Letter, we present albedo measurements for the transiting hot Jupiter HD\,189733b across the wavelength range $\lambda = 290\textnormal{--}570$\,nm.

At these wavelengths, clear atmosphere models (i.e.~without clouds) predict that hot Jupiter albedos are suppressed by alkali absorption \citep{2000ApJ...538..885S, 2008ApJ...682.1277B}. Observations to date have been largely consistent with these expectations \citep[e.g.][]{2008ApJ...689.1345R, 2009A&A...506..353A, 2009Natur.459..543S, 2010AJ....139.1481A, 2010A&A...513A..76S, 2010ApJ...710...97C, 2010ApJ...713L.145W, 2011ApJS..197...11D, 2011MNRAS.417L..88K, 2013ApJ...764L..22M}. Models also predict, however, that silicates and iron could condense in the uppermost layers of some atmospheres, raising the albedo significantly \citep{1999ApJ...513..879M, 2000ApJ...538..885S}. For instance, reflective clouds seem necessary to explain the relatively high albedo of Kepler-7b \citep{2011ApJ...730...50K, 2011ApJ...735L..12D}.  

By measuring the reflection signal of HD\,189733b, our goal was to gauge the role of clouds/hazes\footnote{In this Letter, we use the words ``cloud'' and ``haze'' interchangably as referring to any suspended condensates in the atmosphere.} in the atmosphere of this well-studied hot Jupiter. Motivation came from the observed transmission spectrum, which slopes downward from 290\,nm out to 1\,$\mu$m \citep{2008MNRAS.385..109P, 2011MNRAS.416.1443S} and possibly further into the infrared \citep{2009A&A...505..891S, 2012MNRAS.422..753G}. A likely explanation for this feature is Rayleigh scattering by a high altitude haze of dust \citep{2008A&A...481L..83L}. Furthermore, \cite{2012MNRAS.422.2477H} detected the narrow core of the Na 589\,nm doublet in transmission, but not the pressure-broadened wings that would be expected in a clear atmosphere \citep[e.g.][]{2010ApJ...709.1396F}. Indirect evidence of clouds on the dayside hemisphere has also been identified by \cite{pont_etal_2013}, who speculated that the albedo of HD\,189733b might be high as a result.

In practice, we measured the reflection signal by monitoring the change in brightness of the star-planet system that occurred during secondary eclipse. Unlike the primary transit, which allows us to probe the day-night terminator region of the atmosphere, the secondary eclipse signal is directly related to the brightness of the dayside hemisphere. Although secondary eclipses have been measured previously for HD\,189733b at infrared wavelengths \citep{2006ApJ...644..560D, 2007Natur.447..183K, 2008Natur.456..767G, 2008ApJ...686.1341C, 2010ApJ...721.1861A, 2012ApJ...754...22K}, our new observation constitutes the first measured for HD\,189733b at short wavelengths, where thermal emission from the planet is negligible. Any detected light is therefore entirely due to scattering by the atmosphere. This allows us to place unambiguous constraints on the geometric albedo $A_g$, according to:
\begin{eqnarray}
A_g \, &=& \, \delta \, \left[ \, \rho \, \frac{R_\star}{a} \, \right]^{-2} \ , \label{eq:Ag}
\end{eqnarray}
where $\delta$ is the fractional eclipse depth, $\rho = R_p/R_\star$ is the ratio of the planet and star radii, and $a$ is the orbital semimajor axis. Equation \ref{eq:Ag} follows from the formal definition of $A_g$, namely, the observed flux of scattered light at full phase divided  by the flux that an isotropically scattering disk with the same cross-sectional area would have if it were placed at the same location as the planet \citep[e.g.][]{2010eapp.book.....S}.


\section{ Observations and data reduction } \label{sec:obs_and_datared}

One secondary eclipse of HD\,189733b was observed over four Hubble Space Telescope (\emph{HST}) orbits using the Space Telescope Imaging Spectrograph (STIS) G430L grating (290--570\,nm) for programme GO-13006 (P.I., F.~Pont) on 2012 December 20. Spectra taken during the first orbit exhibited much larger systematics than those taken in subsequent orbits due to settling of the telescope into its new pointing position and were not included in the analysis. Of the remaining three orbits, the first and third sampled the out-of-eclipse flux while the planet was close to full phase (star plus planet), and the second sampled the in-eclipse flux while the planet was fully obscured by the star (star only).

We used a wide $52^{\prime\prime} \times 2^{\prime\prime}$ slit to minimise time-varying slit losses caused by pointing drifts and reduced overheads by reading out only the $1024 \times 128$ pixel subarray containing the target spectrum. Based on previous experience with \emph{HST}/STIS observations, we expected the first exposure of each satellite orbit to have systematically lower counts than the exposures immediately following. For this reason, a dummy 1\,s exposure was taken at the beginning of each orbit, followed by a series of 35 science exposures with integration times of 64\,s. Unfortunately, the first science exposure of each satellite orbit still exhibited systematically lower flux levels, so we discarded these as well. Our final dataset thus consisted of 102 spectra taken over 237\,mins.

\begin{figure}
\centering
\includegraphics[width=\columnwidth]{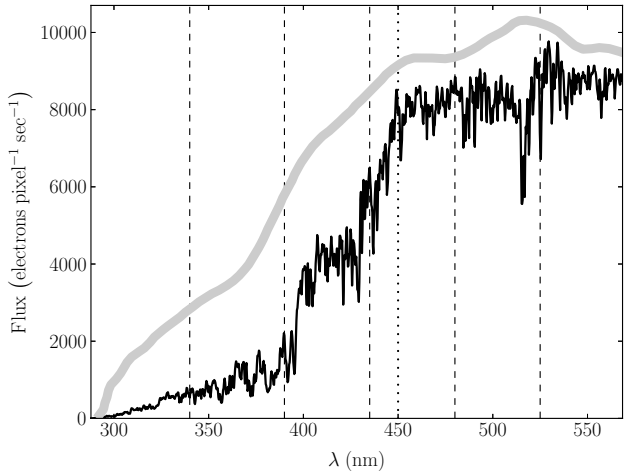}
\caption{ Mean system spectrum constructed from the out-of-eclipse spectra (solid black line) and G430L sensitivity curve with arbitrary normalization (thick gray line). Vertical lines indicate boundaries between the two-channel (dotted line) and six-channel (dashed lines) wavelength bins used for the lightcurve analysis. \label{fig:stellar_spectrum} }
\end{figure}

Images were reduced using the CALSTIS v2.40 pipeline and cleaned for cosmic rays. The spectra were then extracted using the IRAF \verb apall \ routine with a 13 pixel-wide aperture. Background subtraction was not performed, as the background contribution was negligible. Spectra were Doppler-corrected to the heliocentric rest frame, corresponding to shifts of $\sim$1 pixel along the dispersion axis. We generated photometric time series by integrating the flux from each individual exposure across the dispersion axis within different wavelength bins. Separate analyses were performed for a two-channel and six-channel binning: Figure \ref{fig:stellar_spectrum} shows the mean system spectrum with adopted wavelength bins overplotted and Table \ref{tab:reflection_spectrum} explicitly lists the wavelength ranges.

\begin{deluxetable}{ccrr}
\tablecaption{Visible albedo measurements for HD\,189733b \label{tab:reflection_spectrum} }
\tablewidth{0pt}
\tablehead{ 
\colhead{$\Delta \lambda$ (nm)} \hspace{1cm} & \colhead{$\lambda_c$ (nm)}   &  $\delta$ (ppm)& \colhead{$A_g$} 
}
\startdata
\shortlowWav--\shortuppWav & \shortcenWav & $\shortmedDelta^{+\shortuppDelta}_{-\shortlowDelta}$ & $\shortmedAg^{+\shortuppAg}_{-\shortlowAg}$ \medskip \\  
\longlowWav--\longuppWav & \longcenWav & $\longmedDelta^{+\longuppDelta}_{-\longlowDelta}$ & $\longmedAg^{+\longuppAg}_{-\longlowAg}$ \medskip  \\ 
\binAlowWav--\binAuppWav & \binAcenWav & $\binAmedDelta^{+\binAuppDelta}_{-\binAlowDelta}$ & $\binAmedAg^{+\binAuppAg}_{-\binAlowAg}$  \medskip  \\
\binBlowWav--\binBuppWav & \binBcenWav &  $\binBmedDelta^{+\binBuppDelta}_{-\binBlowDelta}$ & $\binBmedAg^{+\binBuppAg}_{-\binBlowAg}$ \medskip \\ 
\binClowWav--\binCuppWav & \binCcenWav &  $\binCmedDelta^{+\binCuppDelta}_{-\binClowDelta}$ & $\binCmedAg^{+\binCuppAg}_{-\binClowAg}$ \medskip \\
\binDlowWav--\binDuppWav & \binDcenWav &  $\binDmedDelta^{+\binDuppDelta}_{-\binDlowDelta}$ & $\binDmedAg^{+\binDuppAg}_{-\binDlowAg}$ \medskip \\
\binElowWav--\binEuppWav & \binEcenWav &  $\binEmedDelta^{+\binEuppDelta}_{-\binElowDelta}$ & $\binEmedAg^{+\binEuppAg}_{-\binElowAg}$ \medskip \\
\binFlowWav--\binFuppWav & \binFcenWav &  $\binFmedDelta^{+\binFuppDelta}_{-\binFlowDelta}$ & $\binFmedAg^{+\binFuppAg}_{-\binFlowAg}$ \enddata
\tablecomments{ $\Delta \lambda$ and $\lambda_c$ are, respectively, the wavelength range and flux-weighted central wavelength for each channel. Uncertainties for $A_g$ have been propagated in quadrature.}
\end{deluxetable}

\section{ Lightcurve analysis } \label{sec:lightcurve_analysis}

\begin{figure*}
\centering
\includegraphics[width=\linewidth]{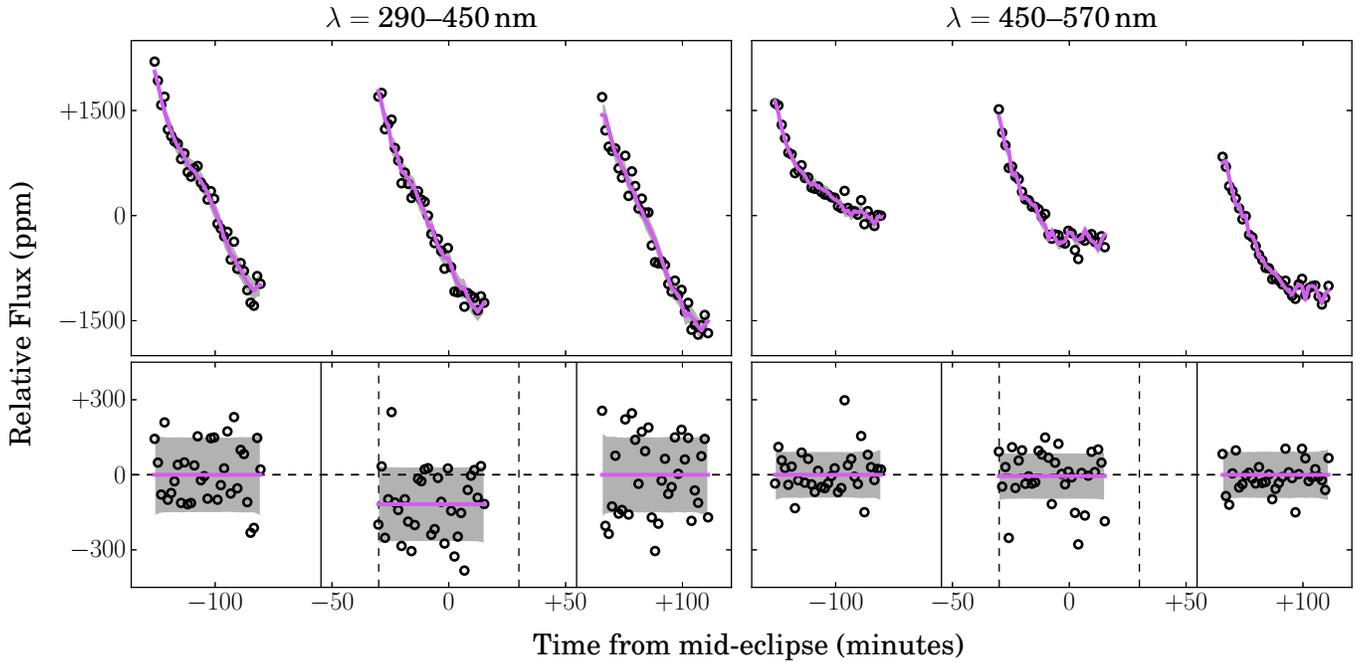}
\caption{Top and bottom panels show, respectively, raw and corrected flux measurements (black circles) with best-fit GP model (purple lines) and per-point $1\sigma$ uncertainties (gray shading). The \shortlowWav--\shortuppWav\,nm and \longlowWav--\longuppWav\,nm wavelength channels are shown on the left and right, respectively. Raw fluxes are median-subtracted and corrected fluxes show the variation about the out-of-eclipse baseline level. Bottom panels also show the start and end of ingress and egress, respectively, (solid vertical lines) and the times of full-eclipse (dashed vertical lines). \label{fig:lightcurves1} }
\end{figure*}

The top panels of Figure \ref{fig:lightcurves1} show the raw lightcurves for the two-channel binning. Within each orbit, the measured flux is dominated by an approximately repeatable decrease of $\sim$\,1500--3000\,ppm. Smaller amplitude correlations on shorter time scales are also evident, as well as a longer term decrease in the baseline flux level. Similar systematics are observed for the six-channel lightcurves, and are believed to be primarily caused by the thermal cycle of the satellite and the drift of the spectral trace across the detector \citep[eg.][]{2001ApJ...552..699B, 2011MNRAS.416.1443S, 2012MNRAS.422.2477H}. 

We modelled the systematics and eclipse signal simultaneously as a Gaussian process (GP), following the approach of \cite{2012MNRAS.422..753G, 2012MNRAS.419.2683G, 2013MNRAS.428.3680G}. Under the GP framework, the model likelihood takes the form of a multivariate normal distribution:
\begin{eqnarray}
p(\mb{f}|\gb{\theta}) &=& \mathcal{N}( \, \gb{\mu}, \mb{K}+\sigma_W^2\mb{I} \, ) \ , \label{eq:gp_logp}
\end{eqnarray}
where $\mb{f} = [ f_1, \ldots, f_N ]^{\rm{T}}$ are the $N=102$ measured fluxes, $\gb{\theta}$ are the model parameters, $\gb{\mu}$ is the model mean vector, $\mb{K}$ is the covariance matrix, $\sigma_W$ is the white noise level, and $\mb{I}$ is the identity matrix. A boxcar function was used for the eclipse signal, such that:
\begin{eqnarray}
\mu_i &=& \alpha \left( 1 - \delta B_i \right) \ , \label{eq:mui}
\end{eqnarray}
where $\alpha$ is the constant baseline flux level, $\delta$ is the fractional flux change during eclipse, and:
\begin{eqnarray}
B_i = \begin{cases}
0 & \rm{during \ 1st \ and \ 3rd \ orbits} \\
1 & \rm{during \ 2nd \ orbit} \ ,
\end{cases} \label{eq:boxcar}
\end{eqnarray}
for $i=1, \ldots, N$.

Off-diagonal entries of the covariance matrix $\mb{K}$ account for correlations between the measured flux and other variables that are unrelated to the eclipse signal, which we refer to as auxiliary variables (see below). We used the squared exponential kernel to populate the entries of $\mb{K}$, such that:
\begin{eqnarray}
K_{ij} &=& C^2 \, \exp \left[ \, - \sum\limits_{d=1}^D{ \left( \frac{ v_{d,i} - v_{d,j}}{L_d} \right)^2} \, \right] \ , \label{eq:sqexp}
\end{eqnarray}
where $C^2$ is the covariance amplitude, $v_{d,i}$ and $v_{d,j}$ are the $i^{\rm{th}}$ and $j^{\rm{th}}$ values of the $d^{\rm{th}}$ auxiliary variable, respectively, and $L_d$ is the correlation length scale of the $d^{\rm{th}}$ auxiliary variable. By parameterizing the covariance according to Equation \ref{eq:sqexp}, we effectively model the poorly understood systematics as varying smoothly with respect to the auxiliary variables, without actually having to specify the functional form. See \cite{2012MNRAS.419.2683G} for more details.

For auxiliary variables, we used the orbital phase of the satellite $\phi$ and the tilt of the spectral trace on the detector $\psi$. The $\phi$-dependence accounted for the dominant systematic that repeated from orbit to orbit, while the $\psi$-dependence accounted for the smaller-amplitude, higher-frequency correlations as well as the longer-term decrease in the flux baseline level (Figure \ref{fig:lightcurves1}). The latter was caused by an overall drift in the value of $\psi$ throughout the observations, in addition to the smaller-amplitude jitter within orbits. We also tried using the measured shifts of the spectral trace along the dispersion $x$-axis and cross-dispersion $y$-axis as additional auxiliary variables, but found their inclusion had an insignificant effect on the result. Our final parameter set therefore consisted of $\gb{\theta}=\{ \, \delta, \, \alpha, \, C, \, L_\phi, \, L_\psi, \, \sigma_W \}$. 

To marginalize Equation \ref{eq:gp_logp} over the space spanned by $\gb{\theta}$, we used the open source software package PyMC \citep{Patiletal2010} to implement Markov Chain Monte Carlo (MCMC) sampling with the Metropolis-Hastings algorithm \citep{MetropolisRosenbluth53, hastings70}. We ran five chains of 120,000 steps each, where a single step consisted of cycling through the parameters and updating their values one at a time. Random step sizes were adjusted separately for each parameter to maintain step acceptance rates of 20--40\% throughout the chains. After discarding the first 20,000 steps as burn-in, the Gelman-Rubin values \citep{GelmanRubin92} were found to be well within 1\% for all parameters, suggesting that the chains had converged and were well-mixed.

As a check, we also modelled the systematics with different linear combinations of the auxiliary variables and used the Bayesian information criterion \citep{Schwarz1978} to choose between models. This has become a standard approach for analyzing \emph{HST}/STIS primary transit lightcurves \citep[e.g.][]{2011MNRAS.416.1443S, 2012MNRAS.422.2477H}. The eclipse depths inferred from these analyses were consistent with those obtained using the GP model, verifying the robustness of the results to the treatment of instrumental systematics.

\section{ Results } \label{sec:results}

Best-fit GP models for the two-channel binning are shown in Figure \ref{fig:lightcurves1}, with corrected lightcurves in the bottom panels. Inferred eclipse depths $\delta$ are reported in Table \ref{tab:reflection_spectrum} for all wavelength channels. The median of the combined MCMC chain is quoted with uncertainties that correspond to ranges either side containing 34\% of the samples. Maximum likelihood estimates were also obtained for each parameter using the Nelder-Mead simplex algorithm \citep{neldermead_1965} to optimize the joint GP likelihood given by Equation \ref{eq:gp_logp} with respect to $\gb{\theta}$, taking the median MCMC chain values as starting points.  In all cases, the binned chain values for $\delta$ had Gaussian-like distributions and the maximum likelihood solutions were very close to the median chain values. 

Table \ref{tab:reflection_spectrum} also lists the values for the geometric albedo $A_g$, calculated using Equation \ref{eq:Ag} with the measured $\delta$ values, $\rho=0.157 \pm 0.001$ \citep{pont_etal_2013}, and $a/R_\star = 8.863 \pm 0.020$ \citep{2010ApJ...721.1861A}. The most striking result is that the measured albedo in the wavelength range \shortlowWav--\shortuppWav\,nm ($A_g = \shortmedAg \pm \shortuppAg$) is significantly higher than it is in the wavelength range \longlowWav--\longuppWav\,nm ($A_g < \longuppAg$). The broad trend of decreasing eclipse depth from shorter to longer wavelengths is also recovered from the six-channel analysis. 

In principle, variations in the brightness of the star itself, rather than the planetary eclipse, could be responsible for the measured signal. This is particularly pertinent for HD\,189733, which is known to be an active K dwarf. However, if we assume 5000K and 4200K NextGen models \citep{1999ApJ...512..377H} for the star and spot spectra, respectively, with solar metallicity and $\log_{10}g = -4.5$, we find the flux drop in the \shortlowWav--\shortuppWav\,nm channel would only be $\sim$\,10\% greater than the flux drop in the \longlowWav--\longuppWav\,nm channel. The measured difference is significantly larger than this. 

We can also estimate the characteristic amplitude of flux variations due to stellar activity using the power law index of $-2.3$ obtained by \cite{2012A&A...539A.137M} for the combined power spectrum of the brightest K dwarfs in the Q1 \emph{Kepler} dataset. Scaling this to the $\sim 1$\% variation amplitude over $\sim 10$\,day timescales appropriate for HD\,189733, we obtain a corresponding amplitude of $\sim$30\,ppm in the 290--450\,nm channel on timescales of 96\,minutes (i.e.~\emph{HST} orbital period). This is less than half the flux change observed and slightly smaller than the uncertainty on $\delta$ due to other sources. We therefore consider it unlikely that stellar variability could account for the signal, and assume that we have indeed measured the planetary eclipse. 

We are not the first to claim that the albedo of HD\,189733b decreases across the visible wavelength range. \cite{2008ApJ...673L..83B, 2011ApJ...728L...6B} used polarimetry to infer albedos of $A_g = 0.61 \pm 0.12$ and $A_g = 0.28 \pm 0.16 $ in the $B$ (390--480\,nm) and $V$ (500--590\,nm) bands, respectively. Our results are systematically $\sim 2\sigma$ lower than these values.

\begin{figure}
\centering
\includegraphics[width=\columnwidth]{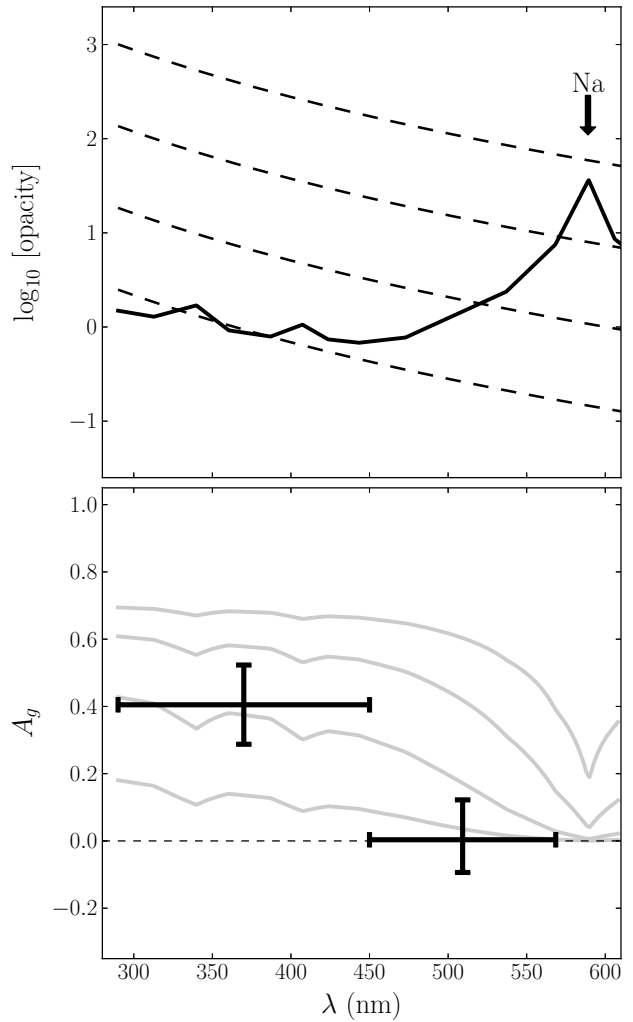}
\caption{\emph{Top panel}: Relative contributions of absorption (solid line) and Rayleigh scattering (dashed lines) to the total opacity in our toy model. The scattering profiles are separated by increments of two atmospheric scale heights, to simulate clouds becoming optically thick at different altitudes. Opacity units are arbitrary as they cancel out in the toy model calculation. For reference, the base of the absorption profile is expected to correspond to pressures of $\sim$1--5\,bar \citep[e.g.][]{2000ApJ...538..885S, 2008ApJ...678.1419F}.  \emph{Bottom panel}: Measured geometric albedos for the two-channel analysis (black crosses), with vertical bars indicating $1\sigma$ uncertainties and horizontal bars indicating the extent of the corresponding wavelength channels. Toy model predictions (gray lines) are also shown for the different cloud altitudes in the top panel.\label{fig:reflection_spectrum} } 
\end{figure}

\section{Implications for the atmosphere} \label{sec:atmosphere_implications}

To first order, the reflection spectrum of a hot Jupiter at visible wavelengths will depend on whether or not there are reflective clouds in the atmosphere, and the altitude at which they become optically thick \citep[e.g.][]{2000ApJ...538..885S}. If scattering clouds are present and become optically thick at altitudes above the absorption wings of the Na 589\,nm doublet and other atomic and molecular absorbers, high albedos ($A_g \sim 0.4\textnormal{--}0.6$) can be maintained throughout the visible wavelength range \citep[e.g.~Class V models of][]{2000ApJ...538..885S}. If there are no clouds, or clouds are present but became optically thick at altitudes well below the absorption, then the albedo can be much lower. Expected values at visible wavelengths in this case will be $A_g < 0.2$ \citep[e.g.~``irradiated'' Class IV models of][]{2000ApJ...538..885S}. In intermediate cases, if clouds are present and become optically thick at pressures comparable to the Na absorption wings, the albedo can be high in the blue channel relative to the red channel\footnote{In the following discussion, we refer to the 290--450\,nm and 450--570\,nm wavelength ranges as the ``blue'' and ``red'' channels, respectively.} \citep[e.g.~``isolated'' Class IV models of][]{2000ApJ...538..885S}.

This list is by no means exhaustive - additional complicating factors could include the possible depletion of Na or the presence of condensates that absorb, rather than scatter, incident light. However, we currently have no reason to suspect that these apply to HD\,189733b, and given the modest resolution and precision of our measurements, we restrict our discussion to the cases described above. Of these, the significantly higher albedo that is measured in the blue channel compared to the red channel is only compatible with the intermediate-altitude cloud scenario. 

To investigate this possibility further, we developed a simple toy model to estimate the expected albedo spectrum. First, we took the photon deposition pressure as a function of wavelength from the pL Class model of \cite{2008ApJ...678.1419F} (see their figure 9) and assumed this pressure was inversely proportional to absorption. Second, we used a Rayleigh $\lambda^{-4}$ scattering profile to model the effect of reflective clouds made up of small dust grains. To simulate clouds that became optically thick at different altitudes, we varied the height of the Rayleigh profile relative to absorption in steps of two atmospheric pressure scale heights, from a high level covering the wings of the Na 589\,nm doublet, down to the altitude where scattering by H$_2$ molecules becomes important. Third, with the absorption and scattering profiles defined, we calculated the albedo according to the two-stream approximation of \cite{2012MNRAS.420...20H}: $A_g = \frac{3}{4}\, (1-\xi^{1/2})/(1+\xi^{1/2})$, where $\xi$ is the ratio of absorption to total opacity (absorption plus scattering) in a given passband. 

The top panel of Figure \ref{fig:reflection_spectrum} shows the adopted opacity profiles, split between absorption and scattering contributions, and the bottom panel shows the predicted albedo spectra with our measured values overplotted. The model with clouds becoming optically thick two scale heights above the base of the absorption profile provides a reasonable fit, although the observations favor a steeper decrease of the albedo toward longer wavelengths. Models with either gray (i.e.~wavelength-independent) scattering and Na absorption, or Rayleigh scattering and gray absorption, were also considered but gave shallower slopes between the two channels.

An interesting question is whether or not the reflection signal is caused by the same scattering species that produces the Rayleigh profile in the transmission spectrum. Given the strong atmospheric circulation expected for hot Jupiters, this seems plausible \citep[e.g.][]{2009ApJ...699..564S, 2012ApJ...751...59P}. \citet{2008A&A...481L..83L} identified enstatite grains (MgSiO$_3$) as likely candidates, being transparent in the visible and formed of atoms abundant in hot atmospheres. 

The limited precision of the current data prevents us from ruling out more intricate scenarios. For example, a high altitude tenuous haze could account for the transmission signal while being transparent at zenith geometry, with a lower, denser layer of clouds producing the reflection signal. The absorption in the red channel could also be due to an absorber other than Na, yet to be identified. For instance, TiO is an efficient absorber in the red channel, although it is expected to have rained out of the atmosphere of HD\,189733b \citep{2008ApJ...678.1419F}.

Our results are suggestive of a low Bond albedo $A_B$, which is defined as the fraction of incident starlight reflected to space at all wavelengths over all angles. This follows from the simple argument that if the Na 589\,nm doublet is not entirely masked by clouds, the situation could be similar at wavelengths beyond 589\,nm where theory predicts significant absorption by the K 770\,nm doublet and molecules such as CH$_4$ and H$_2$O \citep[e.g.][]{2000ApJ...538..885S}. However, our observations do not provide a model-independent constraint on $A_B$, as only $\sim$5\% and $\sim$15\% of the stellar flux is emitted in the blue and red channels, respectively.

\begin{figure}[!t]
\centering
\includegraphics[width=\columnwidth]{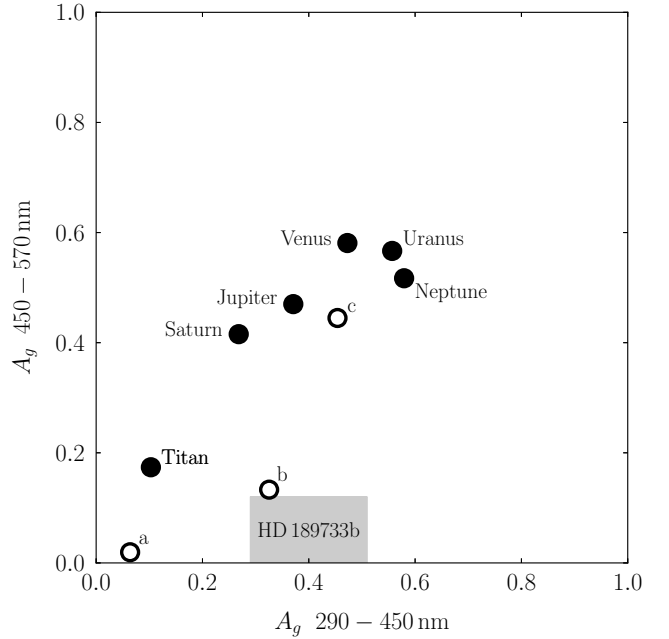}
\caption{Color-color plot showing geometric albedos for various solar system bodies (black circles) with the 1$\sigma$ probability range for HD\,189733b derived in this study (gray box). Solar system albedos are from \cite{1994Icar..111..174K}, except for the Venus albedo which comes from \cite{2007GMS...176..121T}. Also shown are three models from \citet{2000ApJ...538..885S} (open circles): (a) irradiated Class IV, (b) isolated Class IV, and (c) Class V. \label{fig:color_plot} }
\end{figure}

Finally, in Figure \ref{fig:color_plot} we show the blue-versus-red albedos for HD\,189733b and a selection of solar system bodies that also have thick atmospheres, and predicted albedos from three hot Jupiter models. \citet{2011ApJ...728L...6B} hypothesize that HD\,189733b might have a reflection spectrum similar to Neptune. However, our best-fit albedos suggest that it is a deep, dark blue, quite distinct from the atmosphere colors seen in our solar system. 

The results presented here demonstrate the potential of secondary eclipse observations with \emph{HST}/STIS. Our albedo measurements add another piece to the growing jigsaw puzzle of empirical constraints on HD\,189733b's atmosphere, through which we are gradually moving toward a more complete understanding of this exotic planet.

\acknowledgements
Based on observations made with the NASA/ESA Hubble Space Telescope, obtained at the Space Telescope Science Institute, which is operated by the Association of Universities for Research in Astronomy, Inc., under NASA contract NAS 5-26555. 

\clearpage
\bibliographystyle{apj}

\begin{thebibliography}{49}

\bibitem[{{Agol} {et~al.}(2010){Agol}, {Cowan}, {Knutson}, {Deming}, {Steffen},
  {Henry}, \& {Charbonneau}}]{2010ApJ...721.1861A}
{Agol}, E., {Cowan}, N.~B., {Knutson}, H.~A., {Deming}, D., {Steffen}, J.~H.,
  {Henry}, G.~W., \& {Charbonneau}, D. 2010, \apj, 721, 1861

\bibitem[{{Alonso} {et~al.}(2009){Alonso}, {Alapini}, {Aigrain}, {Auvergne},
  {Baglin}, {Barbieri}, {Barge}, {Bonomo}, {Bord{\'e}}, {Bouchy},
  {Chaintreuil}, {de La Reza}, {Deeg}, {Deleuil}, {Dvorak}, {Erikson},
  {Fridlund}, {de Oliveira Fialho}, {Gondoin}, {Guillot}, {Hatzes}, {Jorda},
  {Lammer}, {L{\'e}ger}, {Llebaria}, {Magain}, {Mazeh}, {Moutou}, {Ollivier},
  {P{\"a}tzold}, {Pont}, {Queloz}, {Rauer}, {Rouan}, {Schneider}, \&
  {Wuchterl}}]{2009A&A...506..353A}
{Alonso}, R., {Alapini}, A., {Aigrain}, S., {Auvergne}, M., {Baglin}, A.,
  {Barbieri}, M., {Barge}, P., {Bonomo}, A.~S., {Bord{\'e}}, P., {Bouchy}, F.,
  {Chaintreuil}, S., {de La Reza}, R., {Deeg}, H.~J., {Deleuil}, M., {Dvorak},
  R., {Erikson}, A., {Fridlund}, M., {de Oliveira Fialho}, F., {Gondoin}, P.,
  {Guillot}, T., {Hatzes}, A., {Jorda}, L., {Lammer}, H., {L{\'e}ger}, A.,
  {Llebaria}, A., {Magain}, P., {Mazeh}, T., {Moutou}, C., {Ollivier}, M.,
  {P{\"a}tzold}, M., {Pont}, F., {Queloz}, D., {Rauer}, H., {Rouan}, D.,
  {Schneider}, J., \& {Wuchterl}, G. 2009, \aap, 506, 353

\bibitem[{{Alonso} {et~al.}(2010){Alonso}, {Deeg}, {Kabath}, \&
  {Rabus}}]{2010AJ....139.1481A}
{Alonso}, R., {Deeg}, H.~J., {Kabath}, P., \& {Rabus}, M. 2010, \aj, 139, 1481

\bibitem[{{Berdyugina} {et~al.}(2008){Berdyugina}, {Berdyugin}, {Fluri}, \&
  {Piirola}}]{2008ApJ...673L..83B}
{Berdyugina}, S.~V., {Berdyugin}, A.~V., {Fluri}, D.~M., \& {Piirola}, V. 2008,
  \apjl, 673, L83

\bibitem[{{Berdyugina} {et~al.}(2011){Berdyugina}, {Berdyugin}, {Fluri}, \&
  {Piirola}}]{2011ApJ...728L...6B}
---. 2011, \apjl, 728, L6

\bibitem[{{Brown} {et~al.}(2001){Brown}, {Charbonneau}, {Gilliland}, {Noyes},
  \& {Burrows}}]{2001ApJ...552..699B}
{Brown}, T.~M., {Charbonneau}, D., {Gilliland}, R.~L., {Noyes}, R.~W., \&
  {Burrows}, A. 2001, \apj, 552, 699

\bibitem[{{Burrows} {et~al.}(2008){Burrows}, {Ibgui}, \&
  {Hubeny}}]{2008ApJ...682.1277B}
{Burrows}, A., {Ibgui}, L., \& {Hubeny}, I. 2008, \apj, 682, 1277

\bibitem[{{Charbonneau} {et~al.}(2008){Charbonneau}, {Knutson}, {Barman},
  {Allen}, {Mayor}, {Megeath}, {Queloz}, \& {Udry}}]{2008ApJ...686.1341C}
{Charbonneau}, D., {Knutson}, H.~A., {Barman}, T., {Allen}, L.~E., {Mayor}, M.,
  {Megeath}, S.~T., {Queloz}, D., \& {Udry}, S. 2008, \apj, 686, 1341

\bibitem[{{Christiansen} {et~al.}(2010){Christiansen}, {Ballard},
  {Charbonneau}, {Madhusudhan}, {Seager}, {Holman}, {Wellnitz}, {Deming},
  {A'Hearn}, \& {EPOXI Team}}]{2010ApJ...710...97C}
{Christiansen}, J.~L., {Ballard}, S., {Charbonneau}, D., {Madhusudhan}, N.,
  {Seager}, S., {Holman}, M.~J., {Wellnitz}, D.~D., {Deming}, D., {A'Hearn},
  M.~F., \& {EPOXI Team}. 2010, \apj, 710, 97

\bibitem[{{Deming} {et~al.}(2006){Deming}, {Harrington}, {Seager}, \&
  {Richardson}}]{2006ApJ...644..560D}
{Deming}, D., {Harrington}, J., {Seager}, S., \& {Richardson}, L.~J. 2006,
  \apj, 644, 560

\bibitem[{{Demory} {et~al.}(2011){Demory}, {Seager}, {Madhusudhan}, {Kjeldsen},
  {Christensen-Dalsgaard}, {Gillon}, {Rowe}, {Welsh}, {Adams}, {Dupree},
  {McCarthy}, {Kulesa}, {Borucki}, \& {Koch}}]{2011ApJ...735L..12D}
{Demory}, B.-O., {Seager}, S., {Madhusudhan}, N., {Kjeldsen}, H.,
  {Christensen-Dalsgaard}, J., {Gillon}, M., {Rowe}, J.~F., {Welsh}, W.~F.,
  {Adams}, E.~R., {Dupree}, A., {McCarthy}, D., {Kulesa}, C., {Borucki}, W.~J.,
  \& {Koch}, D.~G. 2011, \apjl, 735, L12

\bibitem[{{D{\'e}sert} {et~al.}(2011){D{\'e}sert}, {Charbonneau}, {Fortney},
  {Madhusudhan}, {Knutson}, {Fressin}, {Deming}, {Borucki}, {Brown},
  {Caldwell}, {Ford}, {Gilliland}, {Latham}, {Marcy}, \&
  {Seager}}]{2011ApJS..197...11D}
{D{\'e}sert}, J.-M., {Charbonneau}, D., {Fortney}, J.~J., {Madhusudhan}, N.,
  {Knutson}, H.~A., {Fressin}, F., {Deming}, D., {Borucki}, W.~J., {Brown},
  T.~M., {Caldwell}, D., {Ford}, E.~B., {Gilliland}, R.~L., {Latham}, D.~W.,
  {Marcy}, G.~W., \& {Seager}, S. 2011, \apjs, 197, 11

\bibitem[{{Fortney} {et~al.}(2008){Fortney}, {Lodders}, {Marley}, \&
  {Freedman}}]{2008ApJ...678.1419F}
{Fortney}, J.~J., {Lodders}, K., {Marley}, M.~S., \& {Freedman}, R.~S. 2008,
  \apj, 678, 1419

\bibitem[{{Fortney} {et~al.}(2010){Fortney}, {Shabram}, {Showman}, {Lian},
  {Freedman}, {Marley}, \& {Lewis}}]{2010ApJ...709.1396F}
{Fortney}, J.~J., {Shabram}, M., {Showman}, A.~P., {Lian}, Y., {Freedman},
  R.~S., {Marley}, M.~S., \& {Lewis}, N.~K. 2010, \apj, 709, 1396

\bibitem[{Gelman \& Rubin(1992)}]{GelmanRubin92}
Gelman, A. \& Rubin, D.~B. 1992, Stat. Sci., 7, 457

\bibitem[{{Gibson} {et~al.}(2013){Gibson}, {Aigrain}, {Barstow}, {Evans},
  {Fletcher}, \& {Irwin}}]{2013MNRAS.428.3680G}
{Gibson}, N.~P., {Aigrain}, S., {Barstow}, J.~K., {Evans}, T.~M., {Fletcher},
  L.~N., \& {Irwin}, P.~G.~J. 2013, \mnras, 428, 3680

\bibitem[{{Gibson} {et~al.}(2012{\natexlab{a}}){Gibson}, {Aigrain}, {Pont},
  {Sing}, {D{\'e}sert}, {Evans}, {Henry}, {Husnoo}, \&
  {Knutson}}]{2012MNRAS.422..753G}
{Gibson}, N.~P., {Aigrain}, S., {Pont}, F., {Sing}, D.~K., {D{\'e}sert}, J.-M.,
  {Evans}, T.~M., {Henry}, G., {Husnoo}, N., \& {Knutson}, H.
  2012{\natexlab{a}}, \mnras, 422, 753

\bibitem[{{Gibson} {et~al.}(2012{\natexlab{b}}){Gibson}, {Aigrain}, {Roberts},
  {Evans}, {Osborne}, \& {Pont}}]{2012MNRAS.419.2683G}
{Gibson}, N.~P., {Aigrain}, S., {Roberts}, S., {Evans}, T.~M., {Osborne}, M.,
  \& {Pont}, F. 2012{\natexlab{b}}, \mnras, 419, 2683

\bibitem[{{Grillmair} {et~al.}(2008){Grillmair}, {Burrows}, {Charbonneau},
  {Armus}, {Stauffer}, {Meadows}, {van Cleve}, {von Braun}, \&
  {Levine}}]{2008Natur.456..767G}
{Grillmair}, C.~J., {Burrows}, A., {Charbonneau}, D., {Armus}, L., {Stauffer},
  J., {Meadows}, V., {van Cleve}, J., {von Braun}, K., \& {Levine}, D. 2008,
  \nat, 456, 767

\bibitem[{Hastings(1970)}]{hastings70}
Hastings, W.~K. 1970, Biometrika, 57, 97

\bibitem[{{Hauschildt} {et~al.}(1999){Hauschildt}, {Allard}, \&
  {Baron}}]{1999ApJ...512..377H}
{Hauschildt}, P.~H., {Allard}, F., \& {Baron}, E. 1999, \apj, 512, 377

\bibitem[{{Heng} {et~al.}(2012){Heng}, {Hayek}, {Pont}, \&
  {Sing}}]{2012MNRAS.420...20H}
{Heng}, K., {Hayek}, W., {Pont}, F., \& {Sing}, D.~K. 2012, \mnras, 420, 20

\bibitem[{{Huitson} {et~al.}(2012){Huitson}, {Sing}, {Vidal-Madjar},
  {Ballester}, {Lecavelier des Etangs}, {D{\'e}sert}, \&
  {Pont}}]{2012MNRAS.422.2477H}
{Huitson}, C.~M., {Sing}, D.~K., {Vidal-Madjar}, A., {Ballester}, G.~E.,
  {Lecavelier des Etangs}, A., {D{\'e}sert}, J.-M., \& {Pont}, F. 2012, \mnras,
  422, 2477

\bibitem[{{Karkoschka}(1994)}]{1994Icar..111..174K}
{Karkoschka}, E. 1994, \icarus, 111, 174

\bibitem[{{Kipping} \& {Bakos}(2011)}]{2011ApJ...730...50K}
{Kipping}, D. \& {Bakos}, G. 2011, \apj, 730, 50

\bibitem[{{Kipping} \& {Spiegel}(2011)}]{2011MNRAS.417L..88K}
{Kipping}, D.~M. \& {Spiegel}, D.~S. 2011, \mnras, 417, L88

\bibitem[{{Knutson} {et~al.}(2007){Knutson}, {Charbonneau}, {Allen}, {Fortney},
  {Agol}, {Cowan}, {Showman}, {Cooper}, \& {Megeath}}]{2007Natur.447..183K}
{Knutson}, H.~A., {Charbonneau}, D., {Allen}, L.~E., {Fortney}, J.~J., {Agol},
  E., {Cowan}, N.~B., {Showman}, A.~P., {Cooper}, C.~S., \& {Megeath}, S.~T.
  2007, \nat, 447, 183

\bibitem[{{Knutson} {et~al.}(2012){Knutson}, {Lewis}, {Fortney}, {Burrows},
  {Showman}, {Cowan}, {Agol}, {Aigrain}, {Charbonneau}, {Deming}, {D{\'e}sert},
  {Henry}, {Langton}, \& {Laughlin}}]{2012ApJ...754...22K}
{Knutson}, H.~A., {Lewis}, N., {Fortney}, J.~J., {Burrows}, A., {Showman},
  A.~P., {Cowan}, N.~B., {Agol}, E., {Aigrain}, S., {Charbonneau}, D.,
  {Deming}, D., {D{\'e}sert}, J.-M., {Henry}, G.~W., {Langton}, J., \&
  {Laughlin}, G. 2012, \apj, 754, 22

\bibitem[{{Lecavelier Des Etangs} {et~al.}(2008){Lecavelier Des Etangs},
  {Pont}, {Vidal-Madjar}, \& {Sing}}]{2008A&A...481L..83L}
{Lecavelier Des Etangs}, A., {Pont}, F., {Vidal-Madjar}, A., \& {Sing}, D.
  2008, \aap, 481, L83

\bibitem[{{Marley} {et~al.}(1999){Marley}, {Gelino}, {Stephens}, {Lunine}, \&
  {Freedman}}]{1999ApJ...513..879M}
{Marley}, M.~S., {Gelino}, C., {Stephens}, D., {Lunine}, J.~I., \& {Freedman},
  R. 1999, \apj, 513, 879

\bibitem[{{McQuillan} {et~al.}(2012){McQuillan}, {Aigrain}, \&
  {Roberts}}]{2012A&A...539A.137M}
{McQuillan}, A., {Aigrain}, S., \& {Roberts}, S. 2012, \aap, 539, A137

\bibitem[{Metropolis {et~al.}(1953)Metropolis, Rosenbluth, Rosenbluth, \&
  Teller}]{MetropolisRosenbluth53}
Metropolis, N., Rosenbluth, A.~W., Rosenbluth, M.~N., \& Teller, A.~H. 1953,
  \jchemphys, 21, 1087

\bibitem[{{Morris} {et~al.}(2013){Morris}, {Mandell}, \&
  {Deming}}]{2013ApJ...764L..22M}
{Morris}, B.~M., {Mandell}, A.~M., \& {Deming}, D. 2013, \apjl, 764, L22

\bibitem[{{Nelder} \& {Mead}(1965)}]{neldermead_1965}
{Nelder}, J.~A. \& {Mead}, R. 1965, Computer Journal, 7, 308

\bibitem[{Patil {et~al.}(2010)Patil, Huard, \& Fonnesbeck}]{Patiletal2010}
Patil, A., Huard, D., \& Fonnesbeck, C.~J. 2010, Journal of Statistical
  Software, 35, 1

\bibitem[{{Perna} {et~al.}(2012){Perna}, {Heng}, \&
  {Pont}}]{2012ApJ...751...59P}
{Perna}, R., {Heng}, K., \& {Pont}, F. 2012, \apj, 751, 59

\bibitem[{{Pont} {et~al.}(2008){Pont}, {Knutson}, {Gilliland}, {Moutou}, \&
  {Charbonneau}}]{2008MNRAS.385..109P}
{Pont}, F., {Knutson}, H., {Gilliland}, R.~L., {Moutou}, C., \& {Charbonneau},
  D. 2008, \mnras, 385, 109

\bibitem[{{Pont} {et~al.}(2013){Pont}, {Sing}, {Gibson}, {Aigrain}, {Henry}, \&
  {Husnoo}}]{pont_etal_2013}
{Pont}, F., {Sing}, D.~K., {Gibson}, N., {Aigrain}, S., {Henry}, G., \&
  {Husnoo}, N. 2013, \mnras, in press

\bibitem[{{Rowe} {et~al.}(2008){Rowe}, {Matthews}, {Seager}, {Miller-Ricci},
  {Sasselov}, {Kuschnig}, {Guenther}, {Moffat}, {Rucinski}, {Walker}, \&
  {Weiss}}]{2008ApJ...689.1345R}
{Rowe}, J.~F., {Matthews}, J.~M., {Seager}, S., {Miller-Ricci}, E., {Sasselov},
  D., {Kuschnig}, R., {Guenther}, D.~B., {Moffat}, A.~F.~J., {Rucinski}, S.~M.,
  {Walker}, G.~A.~H., \& {Weiss}, W.~W. 2008, \apj, 689, 1345

\bibitem[{{Schwarz}(1978)}]{Schwarz1978}
{Schwarz}, G.~E. 1978, \annstat, 6, 461

\bibitem[{{Seager}(2010)}]{2010eapp.book.....S}
{Seager}, S. 2010, {Exoplanet Atmospheres: Physical Processes} ({Princeton:
  Princeton Univ. Press})

\bibitem[{{Showman} {et~al.}(2009){Showman}, {Fortney}, {Lian}, {Marley},
  {Freedman}, {Knutson}, \& {Charbonneau}}]{2009ApJ...699..564S}
{Showman}, A.~P., {Fortney}, J.~J., {Lian}, Y., {Marley}, M.~S., {Freedman},
  R.~S., {Knutson}, H.~A., \& {Charbonneau}, D. 2009, \apj, 699, 564

\bibitem[{{Sing} {et~al.}(2009){Sing}, {D{\'e}sert}, {Lecavelier Des Etangs},
  {Ballester}, {Vidal-Madjar}, {Parmentier}, {Hebrard}, \&
  {Henry}}]{2009A&A...505..891S}
{Sing}, D.~K., {D{\'e}sert}, J.-M., {Lecavelier Des Etangs}, A., {Ballester},
  G.~E., {Vidal-Madjar}, A., {Parmentier}, V., {Hebrard}, G., \& {Henry}, G.~W.
  2009, \aap, 505, 891

\bibitem[{{Sing} {et~al.}(2011){Sing}, {Pont}, {Aigrain}, {Charbonneau},
  {D{\'e}sert}, {Gibson}, {Gilliland}, {Hayek}, {Henry}, {Knutson}, {Lecavelier
  Des Etangs}, {Mazeh}, \& {Shporer}}]{2011MNRAS.416.1443S}
{Sing}, D.~K., {Pont}, F., {Aigrain}, S., {Charbonneau}, D., {D{\'e}sert},
  J.-M., {Gibson}, N., {Gilliland}, R., {Hayek}, W., {Henry}, G., {Knutson},
  H., {Lecavelier Des Etangs}, A., {Mazeh}, T., \& {Shporer}, A. 2011, \mnras,
  416, 1443

\bibitem[{{Snellen} {et~al.}(2009){Snellen}, {de Mooij}, \&
  {Albrecht}}]{2009Natur.459..543S}
{Snellen}, I.~A.~G., {de Mooij}, E.~J.~W., \& {Albrecht}, S. 2009, \nat, 459,
  543

\bibitem[{{Snellen} {et~al.}(2010){Snellen}, {de Mooij}, \&
  {Burrows}}]{2010A&A...513A..76S}
{Snellen}, I.~A.~G., {de Mooij}, E.~J.~W., \& {Burrows}, A. 2010, \aap, 513,
  A76

\bibitem[{{Sudarsky} {et~al.}(2000){Sudarsky}, {Burrows}, \&
  {Pinto}}]{2000ApJ...538..885S}
{Sudarsky}, D., {Burrows}, A., \& {Pinto}, P. 2000, \apj, 538, 885

\bibitem[{{Titov} {et~al.}(2007){Titov}, {Bullock}, {Crisp}, {Renno}, {Taylor},
  \& {Zasova}}]{2007GMS...176..121T}
{Titov}, D.~V., {Bullock}, M.~A., {Crisp}, D., {Renno}, N.~O., {Taylor}, F.~W.,
  \& {Zasova}, L.~V. 2007, Washington DC American Geophysical Union Geophysical
  Monograph Series, 176, 121

\bibitem[{{Welsh} {et~al.}(2010){Welsh}, {Orosz}, {Seager}, {Fortney},
  {Jenkins}, {Rowe}, {Koch}, \& {Borucki}}]{2010ApJ...713L.145W}
{Welsh}, W.~F., {Orosz}, J.~A., {Seager}, S., {Fortney}, J.~J., {Jenkins}, J.,
  {Rowe}, J.~F., {Koch}, D., \& {Borucki}, W.~J. 2010, \apjl, 713, L145

\end{thebibliography}

\end{document}